# LFG: A Generative Network for Real-Time Recommendation

JUNYI LIU

College of Economics and Management, Nanjing University of Aeronautics and Astronautics, Nanjing, 211106, China

junyiliu@nuaa.edu.cn

Chuanmin Mi

College of Economics and Management, Nanjing University of Aeronautics and Astronautics, Nanjing, 211106, China

cmmi@nuaa.edu.cn

Recommender systems are essential information technologies today, and recommendation algorithms combined with deep learning have become a research hotspot in this field. The recommendation model known as LFM (Latent Factor Model), which captures latent features through matrix factorization and gradient descent to fit user preferences, has given rise to various recommendation algorithms that bring new improvements in recommendation accuracy. However, collaborative filtering recommendation models based on LFM lack flexibility and has shortcomings for real-time recommendations, as they need to redo the matrix factorization and retrain using gradient descent when new users arrive. In response to this, this paper innovatively proposes a Latent Factor Generator (LFG) network, and set the movie recommendation as research theme. The LFG dynamically generates user latent factors through deep neural networks without the need for re-factorization or retrain. Experimental results indicate that the LFG recommendation model outperforms traditional matrix factorization algorithms in recommendation accuracy, providing an effective solution to the challenges of real-time recommendations with LFM.

## 1 INTRODUCTION

With the rapid development of the Internet, the number of items for users to choose online is increasing day by day. Consequently, the time cost and trial-and-error cost of user-item interactions are gradually rising, making recommender systems important for user experience and business development [1]. Collaborative filtering, proposed in traditional recommendation models, is the cornerstone of the recommendation field [2], and models combining machine learning continue to use core algorithms such as similarity calculation and matrix factorization. Among them, recommendation algorithm considering latent factors, called LFM (Latent Factor Model) [3]. LFM infers unknown information by capturing the latent feature matrices of users and items, thus SVD and other algorithms based on matrix factorization for recommendation fall under the category of LFM. Despite the inability to provide specific reasons for recommendations, LFM plays a big role due to its superior recommendation performance compared to collaborative filtering algorithms based on similarity calculation [4].

Research on recommender systems is abundant and diversified in algorithms, but many challenges remain [5]. Recommender systems based on matrix factorization have undeniable drawbacks. On one hand, it's challenging to infuse these models with additional information, and the models cannot fit non-linear relationships, thus the learning scope and expressive ability are limited. In view of this, several improvements in recommendation model have been made, extending various SVD variants. Simon Funk and others proposed FunkSVD, which uses gradient descent to optimize the factorized matrix and only calculate the error between the true value at non-missing positions of the original matrix and the predicted value at corresponding positions of the reconstructed matrix. The regularization in FunkSVD is used to prevent overfitting, effectively alleviating the data sparsity problem and improving the accuracy of rating prediction in recommender systems [6]. BiasSVD improves SVD by incorporating user and item biases, considering individual factors of users and items in the recommender system [7], thereby reducing rating prediction errors. SVD++ is an improved algorithm based on BiasSVD, considering additional implicit factors that affect user rating [8], and it has higher prediction accuracy in recommender systems. In summary, the learning ability of SVD and its variants is limited and how to improve their accuracy remains an ongoing area of research. On the other hand, while the improvements to the original recommendation algorithms have enhanced recommendation effects, the more complex model design has also resulted in greater computational overhead. For newly added users or items, the model needs to redo matrix factorization and retrain using gradient descent [9], making it difficult to achieve efficient real-time recommendations.

In response to this, this paper innovatively proposes a Latent Factor Generator (LFG) recommendation model. The LFG model takes the movie latent factor matrix obtained by SVD as a trainable variable, and generates the user latent factors based on previous user ratings and other explicit information, thereby reconstructing the user-item rating matrix. LFG breaks through the limitations of LFM, dynamically generating the user matrix based on previous user ratings without the need to redo matrix factorization or retrain using gradient descent. This paper selects movie recommendation as the research theme, adopts the commonly used MovieLens-100k and MovieLens-1m as the experimental dataset [10], uses RMSE (Root Mean Square Error) as the evaluation metric, and evaluates the recommendation accuracy and real-time recommendation performance of LFG through cross-validation. The experiments show that LFG has higher flexibility in real-time recommendations compared to SVD and BiasSVD, along with higher recommendation accuracy. It also offers high extensibility for integrating additional information. The achievements of this paper enrich the study of recommendation systems. While reducing the computational overhead of real-time recommendation systems, the proposed model offers higher recommendation accuracy, which provides valuable practical evidence and innovative ideas for the combination of real-time recommendations and deep learning.

## 2 PRELIMINARIES

### 2.1 SVD

Singular Value Decomposition (SVD) is one of matrix factorization algorithms, which decomposes a complex matrix into three distinct matrices: one left singular matrix, one diagonal matrix, and the transpose of one right singular matrix. Specifically, given an m-by-n matrix A, SVD can decompose it into three matrices U, S, and $V^T$ as follows:

$$A = USV^T$$

In the formula, U is an orthogonal matrix with m rows and m columns, S is a diagonal matrix with m rows and n columns, with the diagonal values referred to as singular values, and $V^T$ is an orthogonal matrix with n rows and n columns. SVD does not require the decomposed matrix to be a square matrix, and the matrix can be reconstructed to obtain an approximate original matrix. It can be widely applied in various fields such as signal processing, image compression, data dimensionality reduction, and recommender systems, etc.

In the field of recommendation, SVD is one of the mainstream algorithms used to construct the LFM. After the matrix decomposed using SVD, the multiplication of matrices U and $\sqrt{S}$ results in the user latent factor matrix, and the multiplication of $\sqrt{S}$ and $V^T$ results in the item latent factor matrix. Through gradient descent and matrix reconstruction, the rating matrix approximating the original matrix is obtained, based on which users' preference for items with no interaction can be predicted. The singular values obtained from the decomposition are arranged from high to low, and those at the forefront already contain sufficient features. Therefore, usually, the first k singular values are selected, and the matrices are truncated accordingly to calculate the user and item latent factors matrices.

The SVD method is typically only applicable to offline recommendations and cannot meet the requirements for real-time updates. Moreover, the traditional SVD algorithm assumes that the rating preference can be calculated by completing the multiplication user latent factors vector and item latent factor vector and summing up, making it hard to integrate other information or fit non-linear relationships. To address this, modifications to the original SVD algorithm [11] or the combination of SVD with machine learning techniques [12] can complement each other while retaining the advantages of matrix factorization and achieving more flexible and accurate personalized recommendations. In conclusion, SVD is one of the powerful core technologies in recommender system field, possessing the ability to extract latent factors from user and item interaction information [13], and the combination with machine learning is necessary and effective to enable the recommendation model to learn high dimension data features. It is a research direction worth exploring in the modern recommendation field.

### 2.2 Generative Network

Generative network is a type of deep neural network with a strong ability to learn complex and high-dimensional data distributions [14]. They are widely used in various scenarios, including image generation, text summarization, speech synthesis, sample augmentation, and have several variants such as Generative Adversarial Networks (GANs) and Variational Autoencoders (VAEs).

Generative networks possess a strong learning ability and can be used to build highly personalized recommender systems. In the field of recommendation, generative networks can learn user-item interaction information, thoroughly excavate explicit and implicit features of users and items, and learn richer nonlinear relationships between data to achieve accurate rating predictions or generate personalized Top-N recommendation lists.

Furthermore, for the key aspects in the field of recommendation such as information fusion, real-time recommendation, and data sparsity, generative networks also have many advantages. Compared to traditional collaborative filtering or matrix factorization methods, generative networks are adept at integrating multimodal information. They can accept data of different modalities and scales as network input to generate comprehensive and accurate recommendations. Moreover, since generative networks can generate data at any time, they are applicable to scenarios requiring real-time recommendations and have immense potential in dynamic online environments such as streaming services and e-commerce. Additionally, generative networks accept noisy input and have a certain adaptability to sparse data. Existed VAE-based recommendation models have proven in practice their ability to deeply learn the latent feature representations of users and items even when data is incomplete, thereby generating accurate recommendation results [15]. Overall, further exploration and research on generative networks in the field of recommendation are highly valuable and promising.

## 3  PROPOSED METHOD

To break through the limitations of traditional recommendation algorithms and reduce the complexity of integrating machine learning recommendation models, this paper innovatively proposes a Latent Factor Generator network abbreviated as LFG, constructs a flexible and efficient recommendation model, shown in figure 1.

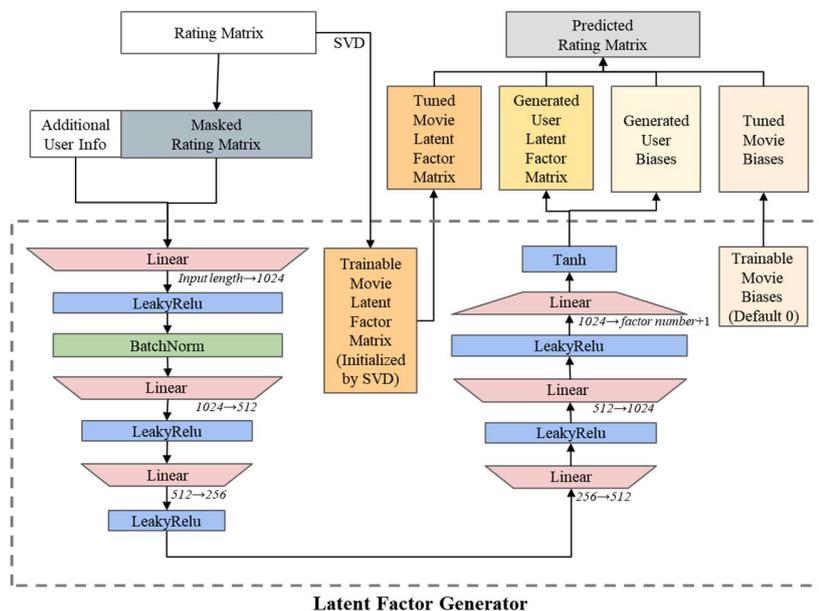

Figure 1: Latent Factor Generator Recommendation Model

## 3.1 The design of Latent Factor Generator Network

The LFG model takes SVD as its upstream, and the number of latent factors corresponds to the truncation value k in SVD, which is set to 50 in this paper. In the results of the SVD, the movie latent factor matrix is set as a trainable variable to be tuned in the LFG model, while the original user latent factor matrix is discarded. To prevent overfitting and enhance the samples, the users' previous m-by-n rating matrix H is randomly masked with a probability of 0.1, concatenated with other auxiliary information matrix E, to be used as the input I of the model:

$$Mask(x,p) = \begin{cases} 0 & with\ probability\ p \\ x & with\ probability\ 1-p \end{cases}$$

$$(H_{masked})_{ij} = Mask(H_{ij}, 0.1) \quad \forall\ i \in \{1, \ldots, m\}, j \in \{1, \ldots, n\}$$

$$I = [E, H_{masked}]$$

The network structure of LFG model mainly consists of Linear layers and LeakyReLU layers. To mitigate the impact of differences between samples on model performance, this paper adopts a Batchnorm layer after the first LeakyReLU layer. Considering that user feedback contains both positive and negative information, the activation function used in the output layer changed to Tanh, which can output negative values compared to Sigmoid or ReLU. The output of the model is the generated user latent factor matrix $U_G$ and the generated user biases $bu_G$, which is used to reconstruct rating matrix $H_{pred}$ with the tuned movie latent factor matrix M, the tuned movie biases $bi_T$ and the average of input ratings μ:

$$H_{pred} = U_G \cdot M + bu_G + bi_T + \mu$$

Though LFG also considers user bias and item bias, its upstream model is simply the original SVD. The initial value of item bias is 0, which is tuned through model updating, and the user bias is generated dynamically by the model. These aspects distinguish it from BiasSVD.

## 3.2 The updating of Latent Factor Generator Network

The training process of LFG uses MSE (Mean Square Error) as the loss function. Specifically, the user ratings are divided into training and test sets, and the rating matrix of training set is regarded as the original. LFG generates user latent factor matrix and user biases based on random-masked original matrix, and through matrix reconstruction, it finally output the predicted rating matrix $H_{pred}$. To provide precise guidance for the update of model parameters, we do not directly calculate the MSE between the original rating matrix and the predicted rating matrix, but only focuses on the error between the true value at non-missing positions of the original matrix and the predicted value at corresponding positions of the reconstructed matrix. The matrix positions in the training set where actual ratings exist are denoted as set T, and the loss function is given by:

$$Loss = \frac{1}{|T|} \sum_{(i,j) \in T} ((H_{pred})_{ij} - H_{ij})^2$$

Through gradient descent, the LFG model update parameters in the right direction, obtaining accurate rating predictions with flexible input.

## 4 EXPERIMENTAL RESULTS

In the field of recommendation, RMSE (Root Mean Square Error) is commonly used as a metric to evaluate model performance:

$$RMSE = \sqrt{\frac{1}{N}\sum_{i=1}^{N}(y_i - \hat{y}_i)^2}$$

In this paper, two experiments are conducted using RMSE as the metric: the basic rating prediction accuracy evaluation and the real-time recommendation accuracy test. Two datasets, MovieLens-100k and MovieLens-1m, were selected to provide a more rigorous assessment of the model's performance [16]. Specifically, the MovieLens-100k dataset contains 100,000 ratings from 943 users on 1682 movies, while the MovieLens-1m dataset comprises 1 million ratings from 6,040 users on 3,883 movies. In both experiments, LFG was compared with SVD and biasSVD, and SVD++ was not included for comparison due to its huge computational overhead.

### 4.1 Basic Rating Prediction Accuracy Evaluation

In this experiment, the previous ratings of each user are proportionally divided for cross-validation, with 80% of each user's ratings allocated to the training set, while the remaining 20% is reserved for evaluation. Additionally, the extra user information participating in the training is constituted by the vectorized user age, gender, and job.

Referring to Table 1 and Table 2, on both the MovieLens-100k dataset and MovieLens-1m dataset, the experimental results show that in each fold of cross-validation, the rating prediction error of the LFG model is consistently lower than that of SVD and BiasSVD, which points out that the LFG model outperforms SVD and BiasSVD in terms of basic rating prediction, especially on larger dataset.

Table 1: LFG Rating Prediction 5-Fold Cross-Validation Results on MovieLens-100k

| Fold | Training Set | Evaluation Set | SVD Eval RMSE | BiasSVD Eval RMSE | LFG Eval RMSE |
|---|---|---|---|---|---|
| 1 | 2,3,4,5 | 1 | 0.9386 | 0.9097 | 0.9085 |
| 2 | 1,3,4,5 | 2 | 0.9393 | 0.9131 | 0.9106 |
| 3 | 1,2,4,5 | 3 | 0.9410 | 0.9160 | 0.9156 |
| 4 | 1,2,3,5 | 4 | 0.9440 | 0.9171 | 0.9175 |
| 5 | 1,2,3,4 | 5 | 0.9392 | 0.9123 | 0.9128 |
| Average | RMSE | | 0.9404 | 0.9136 | 0.9130 |

Table 2: LFG Rating Prediction 5-Fold Cross-Validation Results on MovieLens-1m

| Fold | Training Set | Evaluation Set | SVD Eval RMSE | BiasSVD Eval RMSE | LFG Eval RMSE |
|---|---|---|---|---|---|
| 1 | 2,3,4,5 | 1 | 0.8652 | 0.8594 | 0.8546 |
| 2 | 1,3,4,5 | 2 | 0.8667 | 0.8598 | 0.8554 |
| 3 | 1,2,4,5 | 3 | 0.8686 | 0.8614 | 0.8572 |
| 4 | 1,2,3,5 | 4 | 0.8675 | 0.8610 | 0.8566 |
| 5 | 1,2,3,4 | 5 | 0.8680 | 0.8618 | 0.8591 |
| Average | RMSE | | 0.8672 | 0.8607 | 0.8566 |

## 4.2 Real-Time Recommendation Accuracy Test

Except the basic recommendation accuracy, the performance of real-time recommendation is a key aspect that cannot be overlooked in recommendation models, which is the focus of this paper. The study of real-time recommendation has always been a hot topic of great practical significance in the field of recommendation. To simulate the situation involving new users, this experiment divides the dataset by users and performs SVD again, with 80% of the users utilized for cross-validation and 20% considered as new users for real-time testing.

Recommendation algorithms rooted in SVD lack the capability to accommodate new users. When new data is incorporated into the model, recommendation models based on SVD must redo the matrix factorization to reach high accuracy, but this would cost significant computational overhead. In contrast, the LFG model can directly generate the latent factors including biases for new users based on their information without the need for retraining, resulting in minimal computational overhead. In this experiment, the new users are assigned the global mean value as a placeholder, which ensure their exclusion from training process and also make it possible for subsequent real-time accuracy test. As shown in Table 3 and Table 4, on both datasets the average RMSE values in 5-fold cross-validation indicate the fact that the real-time recommendation accuracy of LFG is significantly superior to SVD and BiasSVD.

Table 3: LFG Real-Time Recommendation 5-Fold Cross-Validation Results on MovieLens-100k

| Fold | Training Set | Evaluation Set | SVD Test RMSE | BiasSVD Test RMSE | LFG Test RMSE |
|---|---|---|---|---|---|
| 1 | 2,3,4,5 | 1 | 1.0979 | 1.0136 | 0.9158 |
| 2 | 1,3,4,5 | 2 | 1.1464 | 1.0460 | 0.9355 |
| 3 | 1,2,4,5 | 3 | 1.1646 | 1.0696 | 0.9416 |
| 4 | 1,2,3,5 | 4 | 1.1057 | 0.9931 | 0.9116 |
| 5 | 1,2,3,4 | 5 | 1.1226 | 1.0303 | 0.9232 |
| | Average | RMSE | 1.1274 | 1.0305 | 0.9255 |

Table 4: LFG Real-Time Recommendation 5-Fold Cross-Validation Results on MovieLens-1m

| Fold | Training Set | Evaluation Set | SVD Test RMSE | BiasSVD Test RMSE | LFG Test RMSE |
|---|---|---|---|---|---|
| 1 | 2,3,4,5 | 1 | 1.0940 | 0.9709 | 0.8789 |
| 2 | 1,3,4,5 | 2 | 1.1218 | 0.9895 | 0.8879 |
| 3 | 1,2,4,5 | 3 | 1.1162 | 0.9763 | 0.8868 |
| 4 | 1,2,3,5 | 4 | 1.1225 | 0.9923 | 0.8801 |
| 5 | 1,2,3,4 | 5 | 1.1208 | 0.9883 | 0.8850 |
| | Average | RMSE | 1.1151 | 0.9835 | 0.8837 |

## 5 DISCUSSION AND CONCLUSION

The essence of the field of recommendation has always been feature engineering, that is, to mine user preferences as much as possible from a small amount or implicit user information, thereby matching users with items that meet their expectations and achieving high recommendation accuracy. Previous studies have made precise and personalized recommender systems by improving traditional algorithms or integrating various machine learning technologies, but when performing real-time recommendations for dynamically growing user information, they always come at the cost of higher model complexity and computational overhead. In view of this, this paper has conducted a meaningful exploration and achieved excellent results.

The focus of this paper is not to improve traditional matrix factorization recommendation algorithms, but to propose an innovative Latent Factor Generator (LFG). The LFG is a deep neural network, regarding matrix factorization as preliminary work. This paper chooses two MovieLens dataset with different scale and conducts sufficient experiments for rating prediction and real-time recommendation. To prevent the randomness of dataset division from affecting model performance and to analyze the stability of the model, all evaluation adopt cross-validation. The results show that the LFG recommendation model proposed in this paper provides an excellent solution for real-time recommendations.

Two experiments are implemented in this paper, and regardless of how the data is processed and divided, both experiments use 80% of the overall information. The first experiment compares the rating prediction accuracy of the LFG model and the SVD algorithm to initially evaluate the model. The results of the 5-fold cross-validation show that the LFG model has higher rating prediction accuracy compared to SVD and BiasSVD. The second experiment is used to test the real-time recommendation performance of the LFG model. To simulate the situation of new users joining, this paper divides users, considering 80% of the users as existed users and the remaining 20% as new users. The information of the new users does not participate in model training in any form. The average RMSE of the LFG model in the real-time recommendation test is significantly lower than SVD and BiasSVD, indicating the LFG model can provide extremely high accuracy for real-time recommendation.

The difference in performance between LFG and SVD is due to their differences in key points of training. LFM based on SVD or its variants focuses on reconstructing the rating information, while LFG focuses on fitting user latent factors. They all minimize the error between predicted ratings and real ratings, but LFG can capture higher-dimensional feature information to generate user latent factor matrix dynamically, achieving flexible real-time recommendations. Furthermore, when predicting based on new data as the second experiment simulated, LFM requires matrix re-factorization or retraining using gradient descent, whereas LFG necessitates only a single inference, leading to a significant reduction in computational cost.

Generative networks have long been applied to the field of recommendation, such as VAE. Both LFG proposed in this paper and VAE learn latent features from user or item information. However, VAE is relatively complex, and LFG cleverly corresponds user's latent features to the latent factors obtained from matrix factorization, making the model concise and effective. In addition, GAN has strong learning ability for complex and high-dimensional data distribution and is theoretically applicable to the field of recommendation. It consists of a generator and a discriminator. Considering that the LFG model aims to minimize the error between predicted ratings and real ratings, MSE can directly provide clear guidance for gradient descent, so LFG does not need any discriminative networks. Although the LFG model is concise and can learn vectorized auxiliary information, it still has certain limitations in the face of multimodal information and other prediction targets in recommender systems. How to integrate LFG with VAE, GAN and others to cope with more complex recommendation scenarios is a subject for further study.

In summary, with movie recommendation as research theme, this paper innovatively proposes the LFG model for recommender system, which shows excellent performance in rating prediction and real-time recommendation considering new users. The work in this paper is a meaningful exploration in the field of recommendation, bringing new ideas to real-time recommendation. Experiments demonstrate that LFG is easy to implement and yields significant results. More realistically, its feature of dynamical generation greatly reduces computational overhead in real-time recommendations. Its potential can be further tapped, and combining it with other recommendation algorithms for application in more recommendation scenarios is our future research work.

# Authors' background

| Your Name | Title* | Research Field | Personal website |
|---|---|---|---|
| **Junyi Liu** (First author) | master student | Deep learning | |
| **Chuanmin Mi** (Corresponding author) | full professor | Management Science and Engineering | http://faculty.nuaa.edu.cn/mcm/en/index.htm |